\documentclass[aip, reprint, twocolumn, amsmath, amsfonts, amssymb]{revtex4-1}

\usepackage{bm}
\usepackage{graphicx}
\usepackage{xcolor}
\usepackage[normalem]{ulem}

\newcommand{\Fock}{\hat F}
\newcommand{\Ham}{\hat H}
\newcommand{\tmu}{\hat\tau_\mu}
\newcommand{\Dmu}{|D_\mu\rangle }
\newcommand{\Tmu}{|T_\mu\rangle }
\newcommand{\wcis}{\omega_{\mathrm{CIS}}}
\newcommand{\wzero}{\omega^{(0)}}

\usepackage[colorlinks=true,linkcolor=blue,citecolor=blue,urlcolor=blue]{hyperref}

\begin{document}

\title{CIS(2): a state-specific, size-intensive perturbative correction
to configuration interaction singles}

\author{Takashi Tsuchimochi}
\email{tsuchimochi@gmail.com}
\affiliation{College of Engineering, Shibaura Institute of Technology,
3-7-5 Toyosu, Koto-ku, Tokyo 135-8548 Japan}

\begin{abstract}
We present a state-specific second-order perturbation theory based on
a configuration interaction singles (CIS) reference and its generalized Fock
operator. Starting from a fully internally contracted
construction, we retain the complete doubles space and internally
contracted triples in the first-order wave function. We consider two partitions, Canonical and Block-diagonal, with
reference excitation energies given by the Fock excitation gap $\wzero$
and the CIS excitation energy $\omega_{\rm CIS}$, respectively.  Our analysis
establishes that strict size-intensivity is preserved when the
contracted-triples denominators use bare Fock gaps and the occupied--virtual
Fock block is projected to remove residual spectator coupling.
Based on this observation, combining Canonical triples with Block-diagonal
doubles yields a size-intensive Hybrid partition.
These properties are verified for water with non-interacting helium atoms.
Our formulation enables $O(o^3v^2)$ cost per matrix--vector product in a
semicanonical basis. 
On the QUEST\#1 benchmark, Canonical and Block-diagonal show opposite
systematic biases, whereas Hybrid gives the best overall accuracy among
the three partitions.  Its mean absolute errors for singlets and
triplets are 0.24 and 0.15~eV, compared with 0.28 and 0.22~eV for
CIS(D).  The improvement is largest for Rydberg excitations, although
CIS(D) remains more accurate for valence singlets.
\end{abstract}

\maketitle

\section{Introduction}
\label{sec:intro}

Accurate excitation energies require a balanced description of electron
correlation in the ground and excited states.  Configuration interaction
singles (CIS)~\cite{ForesmanHeadGordon1992} provides a simple,
size-intensive wave-function description of singly excited states, but its
restricted configuration space omits much of the correlation needed for
quantitative accuracy.  Second-order methods offer a practical compromise
between CIS and more expensive approaches such as equation-of-motion
coupled-cluster theory with single and double excitations
(EOM-CCSD)~\cite{StantonBartlett1993,Izsak2020CCReview}.  In particular,
CIS(D)~\cite{HeadGordon1994CISD,HeadGordon1995CISD} adds a noniterative
correction at fifth-order computational cost, preserving size-intensive excitation energies, 
unlike the earlier CIS-MP2 correction~\cite{ForesmanHeadGordon1992}.
Its direct doubles and
indirect triples contributions substantially improve CIS excitation
energies, and further improvements can be obtained through spin
scaling~\cite{RheeHeadGordon2007,Grimme2007SCSCISD}.
There have also been quasidegenerate extensions, 
which expand the excitation-energy dependence of the doubles
denominator~\cite{HeadGordonOumiMaurice1999CISDn}.

Other widely used second-order approaches include the approximate
coupled-cluster model CC2~\cite{Christiansen1995CC2,Izsak2020CCReview} and the
second-order algebraic diagrammatic construction scheme,
ADC(2)~\cite{Schirmer1982ADC,Dreuw2015ADC}.  These methods obtain excitation energies
from response or propagator equations built on a ground-state reference,
whereas CIS(D) corrects individual CIS states.  Thus, methods of the same
perturbative order can differ in their reference states, amplitude
equations, and treatment of the excitation energy.  The present work
addresses a complementary question: how to define a second-order
correction when the CIS wave function itself is the reference state.

The conventional CIS(D) correction uses Hartree--Fock (HF) orbitals and
M{\o}ller--Plesset denominators.  This construction is effective for many
vertical excitations, but does not directly provide a state-specific
perturbation theory especially for a reference with optimized orbitals.  Such
references are of interest because orbital relaxation can be essential
when the electronic structure changes substantially upon excitation, including charge-transfer states~\cite{Subotnik2011CTBias}.
Several CIS-based approaches incorporate orbital relaxation at different
levels, including variationally orbital-adapted CIS~\cite{LiuSubotnik2014VOACIS},
excited-state mean-field (ESMF)
theory~\cite{SheaNeuscamman2018ESMF,SheaGwinNeuscamman2020,Hardikar2020SCFESMF},
double CIS (DCIS)~\cite{Tsuchimochi2024DCIS}, and state-averaged CIS
(SACIS)~\cite{Tsuchimochi2026SACIS,Tsuchimochi2026SACISgrad}. SACIS also enables the description of conical intersections without specifying an active space.
These approaches nevertheless require additional dynamical correlation
for quantitative excitation energies.  This motivates
a perturbative correction defined in terms of the excited-state reference
itself.

Excited-state M{\o}ller--Plesset perturbation theory (ESMP2) follows this
strategy by correlating an ESMF
reference~\cite{SheaNeuscamman2018ESMF,CluneSheaNeuscamman2020ESMP2}.
Its original $O(N^7)$ cost was reduced to $O(N^5)$ through a modified
zeroth-order Hamiltonian and an approximation to the triples
sector based on natural transition orbitals (NTOs).
Even this fifth-order formulation carries a substantial prefactor because
the leading contractions are repeated in solving the first-order amplitude
equations~\cite{CluneSheaNeuscamman2020ESMP2}. The dependence of this approximation on the NTOs and their response to
nuclear displacements can complicate the formulation of analytical gradients.

These findings motivate reconsidering two elements of CIS-based perturbation theory: the zeroth-order Hamiltonian and the representation
of the first-order wave function.  Multireference perturbation theories
provide useful precedents for both~\cite{Park2020MRReview}.
Complete-active-space second-order perturbation theory
(CASPT2)~\cite{Andersson1990CASPT2,Andersson1992CASPT2} provides an
important precedent for a density-dependent Fock partition, while
$n$-electron valence-state perturbation theory
(NEVPT2)~\cite{Angeli2001NEVPT}, including its strongly and
partially contracted variants~\cite{Angeli2002ContractedNEVPT}, illustrates the role of the partition
and contraction in obtaining separable correlation energies.  Internal
contraction, also used in multireference configuration
interaction~\cite{WernerKnowles1988ICMRCI,Szalay2012MRCIReview}, generates external functions
by applying excitation operators to a reference wave function as a
whole.  CIS provides a particularly simple setting in which to examine
these choices: its reference expansion contains only single substitutions,
so the external doubles and triples can be identified explicitly.

Here we develop a state-specific second-order perturbative correction, denoted CIS(2).
As in ESMP2, we construct the zeroth-order Hamiltonian from the
reference-state generalized Fock operator, here evaluated for the CIS state.
For the first-order wave function, we  start from a fully
internally contracted space, retaining the complete doubles space
and triples contracted with the CIS coefficients. This reduces the number of amplitudes
relative to an uncontracted triples expansion and enables
$O(o^3v^2)$ cost per matrix--vector product, where $o$ and $v$ denote the numbers of occupied and virtual orbitals.
With an appropriate choice of denominators and Fock projection, the
formulation ensures cancellation of the correlation energy of
non-interacting spectator fragments, rigorously guaranteeing
size-intensive excitation energies. 

The choice of zeroth-order Hamiltonian also determines the balance of
correlation between the ground and excited states.  We examine three
partitions, termed Canonical, Block-diagonal, and Hybrid, which differ in
the reference energies used in the doubles and contracted-triples sectors.
Size-intensivity constrains the triples sector but leaves a choice in the
doubles sector.  This distinction motivates Hybrid: it retains the
size-intensive triples denominators while changing the doubles denominators.
As a first step toward extensions to orbital-optimized references, we
develop and assess the present formulation using HF orbitals.

In passing, the name CIS(2) has previously been used by Laikov and
Matsika~\cite{LaikovMatsika2007CIS2} for a quasidegenerate second-order
model in which ground- and excited-state energies are obtained by
diagonalizing a dressed symmetric CI matrix in the reference and singles
space.  The present construction instead corrects an individual CIS state
through a state-specific, internally contracted first-order response.
It does not construct that multistate effective Hamiltonian.

This paper is organized as follows.  Section~\ref{sec:theory} defines the first-order wave
function and partitions, relates the method to CIS(D), and derives the
conditions for size-intensivity and the spin-adapted working equations.
Computational details are given in Sec.~\ref{sec:details}.
Section~\ref{sec:results} first compares CIS(2) with published ESMP2
results on a sixteen-state benchmark, then extends the assessment to
QUEST\#1~\cite{Loos2018QUEST1,Veril2021QUEST}.  It then examines triples contraction, and concludes with numerical size-intensivity tests. Section~\ref{sec:conclusions} summarizes the conclusions.

\section{Theory}
\label{sec:theory}

Throughout this work, $i,j,k,l$ denote occupied orbitals, $a,b,c,d$ virtual orbitals, and $p,q,r,s$ general orbitals. Greek indices $\mu, \nu$ label excitation operators or the corresponding configurations. Amplitude and coupling tensor elements are ordered with virtual indices first, e.g., $u_{abij}$, $t_{abij}$, and $v^{\mathrm D}_{abij}$. We use the Einstein summation convention. 

\subsection{Zeroth-order Hamiltonian and first-order wave function}
\label{sec:h0}

The choice of zeroth-order Hamiltonian determines which interactions
are retained in the reference problem and which generate first-order
amplitudes.  For a CIS reference, this distinction is essential: singles
already enter the reference diagonalization, whereas correlation
requires configurations outside that model space.  We first specify
the partition and then construct the doubles and contracted-triples
representation of the external response.

We take the CIS state as the zeroth-order wave function,
\begin{equation}
|S\rangle  = \sum_{ai} c_{ai}|\Phi_i^a\rangle ,
\label{eq:cis}
\end{equation}
where $|\Phi_0\rangle $ is the HF determinant and $|S\rangle $ is normalized.
The zeroth-order Hamiltonian is built from the CIS-state generalized Fock
operator,
\begin{align}
\Fock &= \sum_{pq} f_{pq}\, a_p^\dagger a_q, \\
f_{pq} &= h_{pq} + \sum_{rs}\Bigl[(pr|qs) - \tfrac12 (ps|qr)\Bigr] D^S_{sr},
\label{eq:genfock}
\end{align}
where $D^S_{pq} = \langle S| a_q^\dagger a_p |S\rangle $ is an element of the CIS one-particle density matrix $\mathbf{D}^S$.  Following the projector construction used in
ESMP2~\cite{SheaNeuscamman2018ESMF}, we define the generalized-Fock partition
\begin{equation}
\hat H_0 = \hat R(\Fock-\Ham)\hat R + \hat P\Ham\hat P + \hat Q\Fock\hat Q,
\label{eq:h0}
\end{equation}
where $\hat R=|S\rangle \langle S|$, $\hat P$ projects onto the reference
determinant and all its single substitutions, and $\hat Q=1-\hat P$.
For HF orbitals, Brillouin's theorem decouples $|\Phi_0\rangle $ from the
singles, so including it in the model space leaves the CIS states unchanged.
The $\hat P\Ham\hat P$ term retains the CIS
eigenvectors, while $\hat R(\Fock-\Ham)\hat R$ changes the eigenvalue
of the target state to its Fock expectation value:
$\hat H_0|S\rangle =E^{(0)}|S\rangle $, with
$E^{(0)}=\langle S|\Fock|S\rangle $.  Outside the model space, the zeroth-order
Hamiltonian is Fock-like.  The corresponding bare Fock excitation gap is
\begin{equation}
\wzero \equiv \mathrm{Tr}[\mathbf{f}\,\bm{\gamma}]
= \langle S|\Fock|S\rangle  - \langle \Phi_0|\Fock|\Phi_0\rangle ,
\label{eq:w0}
\end{equation}
where $\bm{\gamma}=\mathbf{D}^S-\mathbf{D}^{\mathrm{HF}}$ is the CIS
difference density and $\mathbf{D}^{\mathrm{HF}}$ is the HF
one-particle density matrix. Its matrix elements are
\begin{align}
\gamma_{ij}&=-\sum_a c_{ai}c_{aj}, \\
\gamma_{ab}&=\sum_i c_{ai}c_{bi}, \\
\gamma_{ia}&=\gamma_{ai}=0.
\end{align}

We refer to this choice as the Canonical partition.
$E^{(0)}+E^{(1)}=\langle S|\Ham|S\rangle$ recovers the CIS energy
through first order. Alternative choices of the external denominators
are examined in Sec.~\ref{sec:sizeintensive}.

\begin{figure*}[t]
\includegraphics[width=\textwidth]{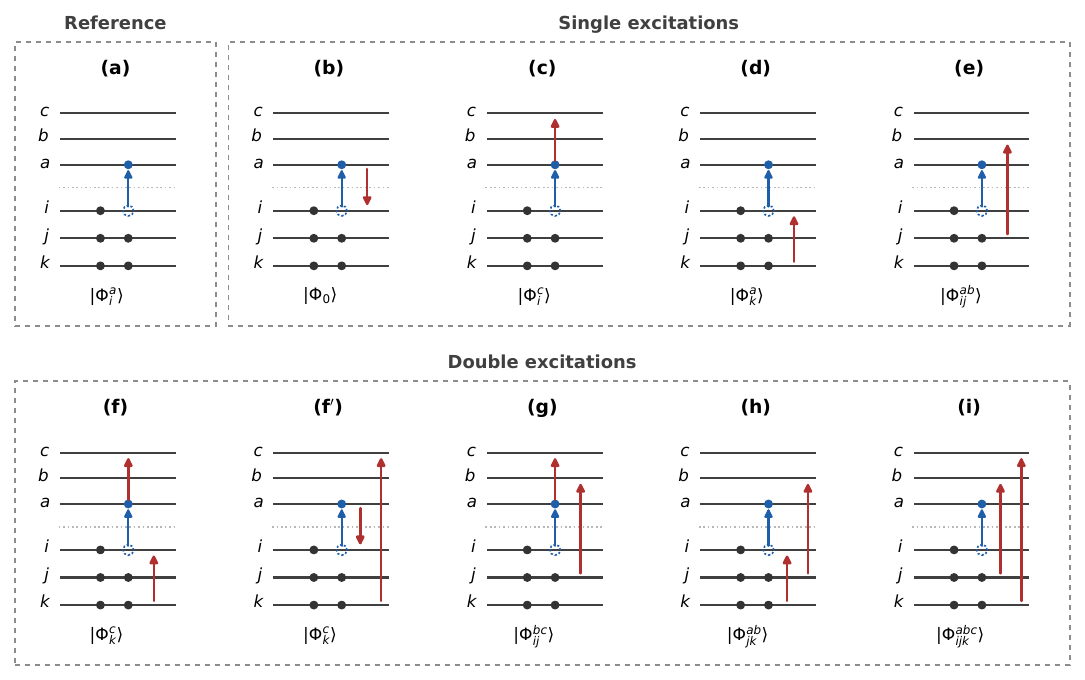}
\caption{Configurations accessible by one- and two-electron excitations of
a CIS determinant.  (a) The initial determinant $|\Phi_i^a\rangle $, with the
promoted electron, hole, and excitation arrow shown in blue.  Dark red arrows
in the remaining panels indicate additional excitations.  Panels (b)--(e)
show one-electron processes and panels (f)--(i) show two-electron processes;
(f) and (f$'$) reach the same determinant by different routes.
The resulting configurations belong to the reference, singles, doubles,
or triples space relative to $|\Phi_0\rangle $.  The panels exhaust the
configurations reachable by one- and two-electron substitutions,
including those containing de-excitations [panel (f$'$)], which
regenerate configurations already shown.  In CIS(2), the triples in (i)
are contracted with the CIS coefficients.}
\label{fig:ansatz}
\end{figure*}

For an isolated target state and with intermediate normalization,
this partition gives a first-order response entirely in the external
space $\hat Q$. To construct our first-order wave function, we
therefore begin with the fully internally contracted space generated
by $\hat Q\hat\Omega|S\rangle$, where $\hat\Omega$ runs over general
one- and two-body operators with unrestricted orbital indices.
Figure~\ref{fig:ansatz} classifies their action on a CIS determinant
by the resulting substitution level relative to $|\Phi_0\rangle$.

The reference and singles in panels (b)--(d), (f), and (f$'$) are
removed by $\hat Q$. Panels (f) and (f$'$) illustrate that different
operator choices can generate the same configuration. The remaining
configurations are doubles [panels (e), (g), and (h)] and triples
[panel (i)]. In the doubles sector, the operators can act on the CIS particle or hole
as well as on other occupied electrons. The resulting functions
contain hole or particle profiles contracted with the CIS
coefficients, so the space they span depends on the rank of
$\mathbf c$. For full-rank $\mathbf c$, they span the complete
doubles space, as shown in the Supporting Information.
Although these functions may be linearly dependent, they can be
replaced by independent double determinants without changing
the first-order space.
In contrast, when the CIS coefficient matrix $\mathbf c$ is rank
deficient, the generated doubles space may be incomplete.
We therefore adopt the more general construction that retains
all double determinants, irrespective of the rank of $\mathbf c$.

The doubles in panels (e), (g), and (h) include configurations
associated with both orbital relaxation and dynamical correlation.
For example, panel (e) retains the original particle--hole pair and
adds a single excitation, as occurs in the first-order response to
an orbital rotation. A change to the doubles denominators can
therefore affect both contributions.

In the triples sector, both additional excitations must involve
other occupied electrons [panel (i)], leaving the CIS particle--hole pair
intact. Its internally contracted functions are therefore
$\tmu|S\rangle$, where $\tmu$ is a $2h$--$2p$ excitation operator
relative to $|\Phi_0\rangle$. The triples component of $\Ham|S\rangle$
lies entirely in this contracted space. Any triple combination
orthogonal to it therefore has zero Hamiltonian coupling to the
CIS reference. Defining
\begin{align}
\Dmu &\equiv \tmu|\Phi_0\rangle
 \qquad (\text{HF-parent double}),\\
\Tmu &\equiv \tmu|S\rangle
 \qquad (\text{contracted triple}),
\end{align}
we obtain the first-order wave function
\begin{equation}
|\Psi^{(1)}\rangle=\sum_\mu u_\mu\Dmu+\sum_\mu t_\mu\Tmu.
\label{eq:firstorder}
\end{equation}
Thus, Eq.~\eqref{eq:firstorder} spans the fully internally
contracted first-order interacting space, augmented where
necessary by additional double determinants.
With $\hat V=\Ham-\hat H_0$, the corresponding second-order energy is
\begin{equation}
E^{(2)}_S=\langle S|\hat V|\Psi^{(1)}\rangle.
\end{equation}

The remaining approximation restricts the triples response
to the internally contracted space, which need not be invariant
under the zeroth-order resolvent. It retains the relative CIS
coefficients for each double excitation operator and requires
$O(o^2v^2)$ triples amplitudes, compared with $O(o^3v^3)$ for an
uncontracted treatment such as the original ESMP2
formulation~\cite{SheaNeuscamman2018ESMF}. The effect of this
restriction on the excitation energies and its role in
size-intensivity are examined in Secs.~\ref{sec:t2c1_valid} and
\ref{sec:sizeintensive}.

\subsection{Working equations}
\label{sec:workeq}

Projecting the first-order perturbation equation onto the doubles
$\langle D_\mu|$ and contracted triples $\langle T_\mu|$ eliminates the
reference-parallel energy term and gives the coupled
linear system
\begin{equation}
\begin{pmatrix}
\mathbf{A}^{\mathrm{DD}} & \mathbf{A}^{\mathrm{DT}}\\[2pt]
\mathbf{A}^{\mathrm{TD}} & \mathbf{A}^{\mathrm{TT}}
\end{pmatrix}
\begin{pmatrix} \mathbf{u} \\[2pt] \mathbf{t} \end{pmatrix}
= -\begin{pmatrix} \mathbf{v}^{\mathrm{D}} \\[2pt] \mathbf{v}^{\mathrm{T}} \end{pmatrix},
\label{eq:linsys}
\end{equation}
with $A^{\mathrm{XY}}_{\mu\nu}=\langle X_\mu|\Fock-E^{(0)}|Y_\nu\rangle $ for
$\mathrm{X},\mathrm{Y}\in\{\mathrm{D},\mathrm{T}\}$ and the perturbation coupling matrix elements
$v^{\mathrm{D}}_\mu=\langle D_\mu|\Ham|S\rangle $,
$v^{\mathrm{T}}_\mu=\langle T_\mu|\Ham|S\rangle $. 
The second-order excitation energy is
\begin{align}
\omega_{\mathrm{CIS(2)}} &= \wcis + E^{(2)}_S - E^{(2)}_0,
\label{eq:e2}\\
E^{(2)}_S &= \tfrac14\sum_{abij}\bigl(u_{abij}v^{\mathrm{D}}_{abij}
 + t_{abij}v^{\mathrm{T}}_{abij}\bigr),
\end{align}
where $E^{(2)}_0$ is the MP2 correlation energy of the ground state. 

We note that the contracted triples $|T_\mu\rangle$ are generally nonorthogonal and can contain linear dependencies, potentially making Eq.~(\ref{eq:linsys}) singular. Null vectors arising from these linear dependencies are orthogonal to the coupling vector ${\bf v}^{\rm T}$ and represent no change in the first-order wave function. Thus, provided that the projected operator is nonsingular on the independent physical subspace, the amplitudes may be nonunique, but the second-order energy $E^{(2)}_S$ is uniquely determined.

\begin{widetext}
\allowdisplaybreaks[1]
Using spin orbitals, antisymmetrized integrals $\langle pq||rs\rangle$,
and the permutation operator
$\mathcal P(pq)X_{pq}=X_{pq}-X_{qp}$, the blocks read:
\begin{align}
(\mathbf{A}^{\mathrm{DD}}\mathbf{u})_{abij} &=
 \mathcal P(ab)\,f_{ac}u_{cbij} - \mathcal P(ij)\,f_{ki}u_{abkj}
 - \omega_{\rm D}\,u_{abij},
\label{eq:add}\\
(\mathbf{A}^{\mathrm{DT}}\mathbf{t})_{abij} &=
 \mathcal P(ab)\mathcal P(ij)\,c_{ai}(\mathbf{t}\mathbf{f})_{bj}
 - \mathcal P(ab)\,(\mathbf{c}\mathbf{f})_{ac}t_{cbij}
 - \mathcal P(ij)\,(\mathbf{f}\mathbf{c})_{ki}t_{abkj}
 + \mathrm{Tr}(\mathbf{f}\mathbf{c})\,t_{abij},
\label{eq:adt}\\
(\mathbf{A}^{\mathrm{TD}}\mathbf{u})_{abij} &=
 \mathcal P(ab)\mathcal P(ij)\,f_{ai}(\mathbf{u}\mathbf{c})_{bj}
 - \mathcal P(ab)\,(\mathbf{c}\mathbf{f})_{ca}u_{cbij}
 - \mathcal P(ij)\,(\mathbf{f}\mathbf{c})_{ik}u_{abkj}
 + \mathrm{Tr}(\mathbf{f}\mathbf{c})\,u_{abij},
\label{eq:atd}
\end{align}
with the intermediates
$(\mathbf{t}\mathbf{f})_{bj}=t_{abij}f_{ia}$, $(\mathbf{u}\mathbf{c})_{bj}=u_{abij}c_{ai}$,
$(\mathbf{c}\mathbf{f})_{ac}=c_{ai}f_{ic}$, and $(\mathbf{f}\mathbf{c})_{ki}=f_{kc}c_{ci}$, and
\begin{align}
(\mathbf{A}^{\mathrm{TT}}\mathbf{t})_{abij} &=
 \mathcal P(ab)\mathcal P(ij)\Bigl[
   -\omega_{\rm T}\,c_{ai}(\mathbf{t}\mathbf{c})_{bj}
   + \bigl((\mathbf{f}\mathbf{c})_{ai}-(\mathbf{c}\mathbf{f})_{ai}\bigr)(\mathbf{t}\mathbf{c})_{bj}
   - c_{ai}(\mathbf{t}\mathbf{c})_{bk}f_{kj} + f_{ac}(\mathbf{t}\mathbf{c})_{ci}c_{bj}
\nonumber\\&\qquad
   + c_{ai}\bigl((\mathbf{f}\mathbf{c})_{ck}-(\mathbf{c}\mathbf{f})_{ck}\bigr)t_{bcjk}
   + f_{ki}\gamma_{ac}t_{cbkj} + f_{ac}\gamma_{ki}t_{cbkj}
 \Bigr]
\nonumber\\&
 + \mathcal P(ab)\Bigl[
   \bigl(\omega_{\rm T}\gamma_{ac} + c_{ak}f_{kl}c_{cl}
   - (\mathbf{f}\bm{\gamma})_{ac} - (\bm{\gamma}\mathbf{f})_{ac} + f_{ac}\bigr)t_{cbij}
   - f_{ac}\gamma_{bd}t_{cdij}\Bigr]
\nonumber\\&
 + \mathcal P(ij)\Bigl[
   \bigl(-\omega_{\rm T}\gamma_{ki} - c_{ck}f_{cd}c_{di}
   - (\mathbf{f}\bm{\gamma})_{ki} - (\bm{\gamma}\mathbf{f})_{ki} - f_{ki}\bigr)t_{abkj}
   - f_{ki}\gamma_{lj}t_{abkl}\Bigr]
 + (\wzero-\omega_{\rm T})t_{abij},
\label{eq:att}
\end{align}
with $(\mathbf{t}\mathbf{c})_{bj}=t_{abij}c_{ai}$. 
Here $\omega_{\rm D}$ and $\omega_{\rm T}$ are the gaps subtracted
in the two sectors, with $\omega_{\rm D}=\omega_{\rm T}=\wzero$
for the Canonical partition of Eq.~\eqref{eq:h0}.
Their independent choice is discussed in Sec.~\ref{sec:sizeintensive}.
The matrix--vector products in Eqs.~\eqref{eq:add}--\eqref{eq:att}
require only the Fock matrix and CIS coefficients;
explicit two-electron integral contractions enter only the
coupling matrix elements,
\begin{align}
v^{\mathrm{D}}_{abij} &= \mathcal P(ij)\,\langle ab||cj\rangle c_{ci}
 - \mathcal P(ab)\,\langle kb||ij\rangle c_{ak}
 + \mathcal P(ij)\mathcal P(ab)\,f^{\rm HF}_{ai}c_{bj},
\label{eq:vd}\\
v^{\mathrm{T}}_{abij} &= \langle ab||ij\rangle 
 + \mathcal P(ij)\mathcal P(ab)\,\langle bc||jk\rangle c_{ai}c_{ck}
 + \mathcal P(ij)\,\gamma_{ki}\langle ab||kj\rangle 
 - \mathcal P(ab)\,\gamma_{ca}\langle cb||ij\rangle.
\label{eq:vt}
\end{align}
\end{widetext}

The amplitudes $\mathbf{u}$ and $\mathbf{t}$ are coupled through
the generally nonzero occupied--virtual block of the CIS-state Fock matrix
in $\mathbf{A}^{\mathrm{DT}}$ and $\mathbf{A}^{\mathrm{TD}}$.
Although this block contributes to neither $E^{(0)}$ nor $E^{(1)}$,
it enters the first-order amplitude equations and affects the second-order
energy substantially.

\subsection{Relation to CIS(D) and ESMP2}
\label{sec:cisd}

CIS(D)~\cite{HeadGordon1994CISD} provides a second-order correction
to the CIS excitation energy but does not define a zeroth-order
Hamiltonian and a corresponding first-order wave function within
a state-specific Rayleigh--Schr\"odinger perturbation framework.
Although an effective first-order wave function can be inferred
from its energy expression, its amplitudes are not obtained as
the solution of a common first-order perturbation equation.
In contrast, CIS(2) explicitly defines both quantities.
To compare the energy expressions, we write the CIS(D) excitation energy as
\begin{align}
\omega_{\mathrm{CIS(D)}} &= \wcis
 - \tfrac14\sum_{abij}\frac{|v^{\mathrm D}_{abij}|^2}{\Delta^{\mathrm{HF}}_{abij}-\wcis}
\nonumber\\
&\quad + \tfrac14\sum_{abij} t^{\mathrm{MP2}}_{abij}\,v^{\mathrm T}_{abij} - E^{(2)}_0 ,
\label{eq:cisd}
\end{align}
where $\Delta^{\mathrm{HF}}_{abij}=\varepsilon_a^{\rm HF}+\varepsilon_b^{\rm HF}
-\varepsilon_i^{\rm HF}-\varepsilon_j^{\rm HF}$ is the HF orbital-energy gap
and $\mathbf{t}^{\mathrm{MP2}}$ represents the ground-state MP2 amplitudes.
The second term is the direct doubles correction.  The last two
terms combine the contracted triples contribution with subtraction of
the ground-state correlation energy.  Eq.~\eqref{eq:cisd} can be
obtained from the CIS(2) energy expression by using the HF Fock matrix,
choosing $\omega_{\rm D}=\wcis$, and fixing the triples amplitudes
at their ground-state MP2 values.

For the HF reference, the occupied--virtual block of the HF Fock matrix
vanishes by the Brillouin condition, giving
$\mathbf A^{\mathrm{DT}}=\mathbf A^{\mathrm{TD}}=0$. In canonical
HF orbitals the doubles block is diagonal, giving
$u_{abij}=-v^{\mathrm D}_{abij}/(\Delta^{\mathrm{HF}}_{abij}-\wcis)$.
The indirect correction is evaluated with
$t^{\mathrm{MP2}}_{abij}=-\langle ab||ij\rangle /\Delta^{\mathrm{HF}}_{abij}$ and the
full contracted coupling vector $\mathbf{v}^{\mathrm T}$.  These amplitudes are imposed rather than obtained by solving the coupled
CIS(2) equations.

CIS(2) replaces this noniterative construction with a state-specific
amplitude problem.  The generalized Fock matrix incorporates the CIS
density, its occupied--virtual block couples the two sectors, and
both $\mathbf u$ and $\mathbf t$ are determined by solving the coupled first-order equations.  The doubles reference energy remains a partition choice.
These changes require an iterative solution, whose cost is discussed in
Sec.~\ref{sec:cost}, but formulate the correction directly in terms of
the CIS reference rather than fixed ground-state MP2 amplitudes.

Although the target reference contains only single substitutions, a
double excitation of electrons outside the excited particle--hole pair
is a triple relative to $|\Phi_0\rangle $.  Such configurations recover
correlation that is also present in the ground state.  Omitting them
would upset the balance in Eq.~\eqref{eq:e2}, even for an excitation
localized far from the correlated electron pair.  CIS(D) supplies this
contribution through ground-state MP2 amplitudes; CIS(2) instead solves
for amplitudes in the contracted space.

ESMP2 provides the closest state-specific comparison.  Its original
formulation uses the same form of the zeroth-order Hamiltonian, but employs an orbital-relaxed reference and an uncontracted
external space~\cite{SheaNeuscamman2018ESMF}.  Clune and
co-workers subsequently reduced its scaling to $O(o^2v^3)$ by
transforming to transition-orbital pairs and simplifying the
zeroth-order Hamiltonian in selected triples
subspaces~\cite{CluneSheaNeuscamman2020ESMP2}.  CIS(2) takes a
different route: it contracts the triples with the CIS coefficients
and retains the projected Fock matrix elements in that space.
We show below that, with HF orbitals, contraction changes the excitation energies only modestly, while substantially reducing the
computational cost.

\subsection{Size-intensivity and partition choices}
\label{sec:sizeintensive}

Size-intensivity requires that an excitation localized on a chromophore $A$
remain unchanged when a non-interacting spectator fragment $B$ is added.
Otherwise, an apparent change in the excitation energy could arise simply
from enlarging the system, even in the absence of a physical interaction.

For non-interacting fragments, $\Ham=\hat H_A+\hat H_B$, CIS configurations
with the excitation on $A$ do not couple to those with the excitation on
$B$, so CIS itself is size-intensive and a CIS state localized on $A$
factorizes as $|S\rangle=|S(A)\rangle\otimes|\Phi_0(B)\rangle$, where
$|\Phi_0(B)\rangle$ is the HF determinant of the spectator fragment.
We examine the conditions under which CIS(2) preserves this
size-intensivity.  The double excitation operators fall into four
classes according to their fragment indices: confined to $A$, confined
to $B$ (spectators), mixed (one single excitation on each fragment), and
charge transfer.  Only the correlation correction of $A$ should remain
in the excitation energy after subtraction of the ground-state
correlation energy.

The mixed operators generate configurations of the form
$|X(A)\rangle\otimes|\Phi_j^b(B)\rangle$ in either sector, with
$|X(A)\rangle=|\Phi_i^a(A)\rangle$ for doubles and
$|X(A)\rangle=\hat a_a^\dagger\hat a_i|S(A)\rangle$ for contracted triples.
Their Hamiltonian coupling to $|S\rangle$ vanishes because
$\langle\Phi_0(B)|\Phi_j^b(B)\rangle=0$ and
$\langle\Phi_0(B)|\hat H_B|\Phi_j^b(B)\rangle=0$, the latter by
Brillouin's theorem.  Charge-transfer configurations likewise have no Hamiltonian coupling
to the CIS reference, since $\hat H_A+\hat H_B$ conserves the electron number on each
fragment.  Because $E^{(2)}_S$ in Eq.~\eqref{eq:e2} contracts the
amplitudes with the Hamiltonian couplings to the CIS reference, mixed
and charge-transfer configurations can affect the energy only indirectly
through the amplitude equations.  In the
non-interacting limit, however, the generalized Fock matrix has no
matrix elements between orbitals on different fragments, and its
occupied--virtual block vanishes on $B$, where the CIS-state Fock
matrix equals the HF Fock matrix of $B$ and is diagonal in its
canonical orbitals.  Every nonzero element of $\mathbf f_{ov}$ thus has
both indices on $A$, where the CIS vector $\mathbf c$ also resides.
Consequently, every block of Eq.~\eqref{eq:linsys} preserves the
fragment assignment, and the only intersector Fock coupling that
survives between spectator-local configurations is the trace coupling
identified below.  The mixed and charge-transfer configurations therefore decouple and may
be dropped.

For a double excitation confined to $B$, the two parents give
\begin{align}
|D_B\rangle&=|\Phi_0(A)\rangle\otimes\hat\tau_\mu(B)|\Phi_0(B)\rangle,\\
|T_B\rangle&=|S(A)\rangle\otimes\hat\tau_\mu(B)|\Phi_0(B)\rangle.
\end{align}
Their Hamiltonian couplings to the CIS reference are $v_B^{\rm D}=0$ and
$v_B^{\rm T}=V_B\equiv\langle ab||ij\rangle$, respectively.
On $B$, the CIS-state Fock matrix equals the HF Fock matrix.  Thus,
for $a,b,i,j\in B$, the triples diagonal element is
$A_{BB}^{\rm TT}=\Delta_B+\wzero-\omega_{\rm T}$, where
$\Delta_B=\varepsilon_a^{\rm HF}+\varepsilon_b^{\rm HF}
-\varepsilon_i^{\rm HF}-\varepsilon_j^{\rm HF}$.
Therefore, size-intensivity requires (1) $\omega_{\rm T}=\wzero$
and (2) no coupling between $|D_B\rangle$ and $|T_B\rangle$.
When these conditions are satisfied, we have for $a,b,i,j\in B$
\begin{align}
 u_{abij}&= 0,\\
 t_{abij}&=-\frac{\langle ab||ij\rangle}{\Delta_B}
\label{eq:spectator_draft}
\end{align}
and the second-order correlation
energy separates as $E^{(2)}_S=E^{(2)}_{S(A)}+E^{(2)}_{{\rm MP2}(B)}$,
where $E^{(2)}_{{\rm MP2}(B)}$ is the MP2 correlation energy of the spectator.
This contribution cancels upon subtraction of the ground-state
correlation energy.

Given that $a,i \in A$ for nonzero $f_{ia}$ and $c_{ai}$, the last terms in
Eqs.~\eqref{eq:adt} and \eqref{eq:atd} couple
the spectators $|D_B\rangle$ and $|T_B\rangle$ through
$g_A={\rm Tr}({\bf fc})$.
For states belonging to a non-totally-symmetric irreducible representation,
$g_A$ vanishes by symmetry.
For totally symmetric states, however, it is generally nonzero.
We therefore remove this coupling by projecting the CIS component out of the
occupied--virtual Fock block,
\begin{equation}
f_{ia}^{\perp}=f_{ia}-g_Ac_{ai},
\label{eq:fperp}
\end{equation}
which ensures $\mathrm{Tr}(\mathbf f_{ov}^{\perp}\mathbf c)=0$.
The virtual--occupied block is its Hermitian transpose.
The projection leaves the occupied--occupied and virtual--virtual
blocks, $\wzero$, and the coupling vectors unchanged. Both size-intensivity conditions (1) and (2) can therefore be satisfied
simultaneously. Although the numerical effect of the projection is small for the
states studied below, it is required to ensure this cancellation
(Sec.~\ref{sec:si_test}). 

The Canonical partition satisfies the cancellation condition with
$(\omega_{\rm D},\omega_{\rm T})=(\wzero,\wzero)$.
An alternative Block-diagonal partition retains the CIS energy in the
model space,
\begin{align}
\hat H_0^{\rm BD}&=\hat P\Ham\hat P+\hat Q\bar F^{\perp}\hat Q,\\
\bar F^{\perp}&=\Fock^{\perp}
+\bigl(E_{\rm HF}-\langle\Phi_0|\Fock^{\perp}|\Phi_0\rangle\bigr)\hat I,
\end{align}
where $\Fock^{\perp}$ denotes the one-body operator with the projected
occupied--virtual blocks.  Its external gaps are referenced to $\wcis$,
giving $(\omega_{\rm D},\omega_{\rm T})=(\wcis,\wcis)$.
The residual chromophore-dependent term $\wzero-\wcis$ in the spectator
triples denominator prevents size-intensivity even after Fock
projection, so the excitation energy drifts with the number of
spectators.

The doubles gap, however, remains unconstrained because the homogeneous
spectator doubles equation admits $\mathbf u_B=\mathbf 0$ for any
$\omega_{\rm D}$. Choosing $\omega_{\rm D}=\wcis$ while retaining
$\omega_{\rm T}=\wzero$ therefore defines a size-intensive Hybrid
partition.
The three partitions are then distinguished by the pair
$(\omega_{\rm D},\omega_{\rm T})$,
\begin{equation}
(\omega_{\rm D},\omega_{\rm T}) =
\begin{cases}
(\wzero, \wzero), & \text{Canonical (size-intensive)},\\
(\wcis, \wcis),   & \text{Block-diagonal (\emph{not})},\\
(\wcis, \wzero),  & \text{Hybrid (size-intensive)} .
\end{cases}
\end{equation}
Relative to Canonical, Hybrid changes only the doubles block of the
working equations by $(\wzero-\wcis)\mathbf I$; the triples and coupling
blocks and both coupling vectors remain unchanged, and the coupled solution lets
the triples amplitudes respond only indirectly.  When $\wzero>\wcis$, the doubles denominators increase.
For the states studied here, this change reduces the negative
bias observed for Canonical (Sec.~\ref{sec:quest}).  Hybrid also uses the
same subtracted doubles gap as CIS(D) (see Eq.~\eqref{eq:cisd}), although its generalized Fock
matrix and solved triples amplitudes make it a distinct approximation.

Finally, we consider the uncontracted $U_3$ ansatz,
\begin{equation}
|\Psi^{(1)\prime}\rangle
= \frac14\sum_{abij}u_{abij}|\Phi_{ij}^{ab}\rangle
+ \frac1{36}\sum_{abcijk}u_{abcijk}|\Phi_{ijk}^{abc}\rangle.
\end{equation}
We denote the full doubles and triples spaces relative to
$|\Phi_0\rangle$ by $U_2$ and $U_3$, respectively. The doubles space
$U_2$ is unchanged from CIS(2).
This ansatz fails to preserve size-intensivity under the Canonical partition
for a reason related to the Block-diagonal case.
Its spectator triples are not contracted against $|S\rangle$,
and their denominators retain a chromophore-dependent contribution,
preventing recovery of the spectator MP2 correlation energy.
As a result, the size-intensivity error is $O(1)$ per spectator.

We will demonstrate these results numerically in  Sec.~\ref{sec:si_test}.

\subsection{Semicanonical orbitals and computational cost}
\label{sec:cost}

A direct evaluation of the CIS(2) matrix--vector product contains terms
scaling as $O(o^3v^2+o^2v^3)$. Since CIS(2) is invariant under separate rotations within the occupied
and virtual orbital spaces, diagonalizing the occupied and virtual
blocks of the CIS-state generalized Fock matrix $\mathbf f$ gives
semicanonical eigenvalues $\varepsilon_p$ and simplifies these
contractions. For example, the doubles block becomes
\begin{equation}
(\mathbf{A}^{\mathrm{DD}}\mathbf{u})_{abij}
=(\Delta_{abij}-\omega_{\rm D})u_{abij},
\end{equation}
where
\begin{equation}
\Delta_{abij}=\varepsilon_a+\varepsilon_b-\varepsilon_i-\varepsilon_j.
\end{equation}
The doubles matrix--vector product therefore requires only
$O(o^2v^2)$ operations.

Contractions over the virtual indices of $\bm{\gamma}$ can also be
factorized without approximation by using $\gamma_{ac}=c_{al}c_{cl}$.
For example, one term in Eq.~\eqref{eq:att} becomes
\begin{equation}
f_{ki}\gamma_{ac}t_{cbkj}
=\varepsilon_i c_{al}c_{cl}t_{cbij}.
\end{equation}
Instead of contracting $\gamma_{ac}$ directly with $t_{cbij}$ at
$O(o^2v^3)$ cost, we first form $w_{lbij}=c_{cl}t_{cbij}$
 and then evaluate $\varepsilon_i c_{al}w_{lbij}$.
Each of the two contractions costs $O(o^3v^2)$. Since typically $o \ll v$, this factorization substantially reduces the computational cost.
The same factorization strategy applies to contractions involving
$(\mathbf c\mathbf f)_{ab}=c_{ak}f_{kb}$. In the spin-orbital implementation, factorization and fusion of
contractions reduce the matrix--vector product to ten
$O(o^3v^2)$ operations.

The coupling vectors $\mathbf v^{\mathrm D}$ and
$\mathbf v^{\mathrm T}$ are formed once and contain intermediates
also used in CIS(D); their construction can benefit from
resolution-of-the-identity techniques~\cite{Hattig2000RICC2}.  Therefore, the iterative solution is the principal additional cost
relative to CIS(D). An uncontracted $U_3$ treatment requires
$O(o^4v^3+o^3v^4)$ work in a general basis and $O(o^3v^3)$ in a
semicanonical basis for the uncontracted implementation considered here,
one power of system size higher than the contracted formulation.

The low-rank structure of the CIS transition
density~\cite{Martin2003NTO} offers an opportunity for further
factorization of the amplitude contractions. Retaining all nonzero
singular values preserves the original equations, whereas truncation
introduces an additional approximation. However, fewer arithmetic
operations need not yield a proportional reduction in elapsed time,
because the full $O(o^2v^2)$ amplitude tensors must still be read and
written. This optimization is therefore not pursued in the present
assessment.

\subsection{Spin-adapted working equations}
\label{sec:impl}

For a closed-shell reference, the amplitude equations can be evaluated
using spatial tensors without constructing their spin-orbital counterparts.
In practice, spin adaptation reduces the amplitudes to spatial tensors,
lowering the computational cost.

For the contracted triples, it must account for the spin of the
CIS reference as well as that of the excitation operator.
Spin-flip symmetry and removal of the quintet component from the
contracted triples determine the independent amplitudes and their index
symmetries.  A singlet requires two pair-symmetric tensors of dimension
$o^2v^2$, one for each excitation sector.  A triplet requires three
tensors, one of which contains two symmetry sectors.  The triplet
matrix elements are related to the singlet expressions by sign changes,
allowing both spin cases to use the same contraction routines.

The explicit working equations, spin adaptation, and packed storage
are described in Sec.~S2 of the Supporting Information.

\section{Computational details}
\label{sec:details}

All calculations were performed using our in-house Python
implementation based on PySCF~\cite{Sun2020PySCF}.
The CIS reference states were obtained by diagonalizing the CIS
Hamiltonian in restricted HF orbitals. The CIS(2) amplitude equations
were solved using right-preconditioned restarted
GMRES~\cite{SaadSchultz1986GMRES}, with a relative-residual threshold
of $10^{-5}$ unless otherwise noted.
The Fock projection of Eq.~\eqref{eq:fperp} is applied to every Canonical and
Hybrid result reported below. 

The first benchmark comprises
the sixteen singlet excitations of Ref.~\onlinecite{SheaGwinNeuscamman2020}; the two
intermolecular charge-transfer cases in that study are not included in
this subset.  These excitations are computed here in the
cc-pVDZ basis~\cite{Dunning1989Basis,Woon1993Basis} with frozen cores;
EOM-CCSD~\cite{StantonBartlett1993} and
CC2~\cite{Christiansen1995CC2} values were computed with PySCF for the same
states.  ESMP2 errors are taken from Ref.~\onlinecite{SheaGwinNeuscamman2020} and aggregated over the same
sixteen states against the same $\delta$-CR-EOM-CC(2,3),D~\cite{Wloch2006CREOMCC,Piecuch2009CREOMCC}
reference values.
The published ESMP2 calculations use frozen cores and cc-pVDZ with an
orbital-optimized ESMF reference; the molecular geometries are taken from
that work's Supporting Information.

\begin{table*}[t]
\caption{Reference excitation energies and signed errors (calculated minus
reference), in eV, for the sixteen singlet excitations
of Ref.~\onlinecite{SheaGwinNeuscamman2020}, using cc-pVDZ and frozen cores.
The Reference column lists the $\delta$-CR-EOM-CC(2,3),D excitation energies;
all method columns report errors. }
\label{tab:table2}
\begin{ruledtabular}
\begingroup
\small
\setlength{\tabcolsep}{3pt}
% generated by gen_tables.py from bench_esmf_table2_eommatch/table2_clean_stats.csv
% ESMP2: esmp2_literature_16.csv, verified against DOI 10.1021/acs.jctc.9b01105.s001, Table 2
\begin{tabular}{lccccccccc}
Molecule & Reference & CIS & CIS(D) & Canonical & Block-diag. & Hybrid & EOM-CCSD & CC2 & ESMP2 \\
\hline
Acetaldehyde & 4.36 & $+0.72$ & $+0.25$ & $-0.30$ & $+0.10$ & $-0.12$ & $+0.21$ & $+0.34$ & $+0.15$ \\
Ammonia & 7.57 & $+0.95$ & $+0.06$ & $-0.05$ & $+0.25$ & $+0.14$ & $+0.05$ & $+0.05$ & $-0.01$ \\
Carbon Monoxide & 8.76 & $+0.60$ & $+0.51$ & $-0.14$ & $-0.10$ & $-0.12$ & $+0.29$ & $+0.36$ & $-0.16$ \\
Cyclopropene & 7.97 & $+0.57$ & $-0.33$ & $-0.27$ & $+0.02$ & $-0.14$ & $-0.08$ & $-0.19$ & $-0.06$ \\
Diazomethane & 3.01 & $+0.39$ & $+0.82$ & $+0.20$ & $+0.53$ & $+0.35$ & $+0.46$ & $+0.62$ & $-0.04$ \\
Dinitrogen & 10.36 & $-1.31$ & $+0.62$ & $+0.10$ & $+0.70$ & $+0.30$ & $+0.44$ & $+0.63$ & $+0.06$ \\
Ethylene & 8.80 & $-0.25$ & $+0.12$ & $-0.30$ & $-0.21$ & $-0.27$ & $+0.19$ & $+0.27$ & $-0.29$ \\
Formaldehyde & 4.08 & $+0.63$ & $+0.25$ & $-0.27$ & $+0.00$ & $-0.11$ & $+0.19$ & $+0.32$ & $+0.15$ \\
Formamide & 5.86 & $+0.88$ & $+0.17$ & $-0.25$ & $+0.27$ & $-0.01$ & $+0.21$ & $+0.34$ & $+0.13$ \\
Hydrogen Sulfide & 7.05 & $+0.59$ & $+0.25$ & $+0.01$ & $+0.16$ & $+0.09$ & $+0.11$ & $+0.22$ & $-0.07$ \\
Ketene & 3.78 & $+0.71$ & $+0.60$ & $-0.02$ & $+0.34$ & $+0.15$ & $+0.36$ & $+0.58$ & $-0.05$ \\
Methanimine & 5.35 & $+0.66$ & $+0.35$ & $-0.21$ & $+0.03$ & $-0.08$ & $+0.22$ & $+0.33$ & $-0.02$ \\
Nitrosomethane & 1.85 & $+0.27$ & $+0.30$ & $-0.43$ & $-0.24$ & $-0.35$ & $+0.25$ & $+0.26$ & $+0.17$ \\
Streptocyanine Cation & 7.53 & $+1.55$ & $+0.07$ & $-0.98$ & $-0.37$ & $-0.62$ & $+0.28$ & $+0.31$ & $-0.42$ \\
Thioformaldehyde & 2.18 & $+0.58$ & $+0.27$ & $-0.16$ & $+0.11$ & $-0.01$ & $+0.24$ & $+0.31$ & $-0.08$ \\
Water & 8.30 & $+1.02$ & $-0.12$ & $-0.10$ & $+0.22$ & $+0.11$ & $-0.02$ & $-0.11$ & $+0.05$ \\
\hline
ME & --- & $+0.53$ & $+0.26$ & $-0.20$ & $+0.11$ & $-0.04$ & $+0.21$ & $+0.29$ & $-0.03$ \\
MAE & --- & 0.73 & 0.32 & 0.24 & 0.23 & 0.19 & 0.22 & 0.33 & 0.12 \\
Max.\ abs.\ error & --- & 1.55 & 0.82 & 0.98 & 0.70 & 0.62 & 0.46 & 0.63 & 0.42 \\
\end{tabular}
\endgroup
\end{ruledtabular}
\end{table*}

The QUEST\#1 benchmark~\cite{Loos2018QUEST1} comprises the vertical excitations of
eighteen small molecules with the geometries, the aug-cc-pVTZ basis
set~\cite{Dunning1989Basis,Kendall1992Aug,Woon1993Basis} and the
frozen-core convention of the original work.  The theoretical best estimates (TBE)
of Ref.~\onlinecite{Loos2018QUEST1} serve as reference.
The class-resolved statistics include 25 valence and 27 Rydberg singlets,
and 29 valence and 18 Rydberg triplets.  The remaining singlet
(hydrogen chloride), which has charge-transfer character,
is included in the overall statistics only.

The size-intensivity and contraction tests use water and formaldehyde
at their QUEST\#1 CC3/aug-cc-pVTZ geometries~\cite{Loos2018QUEST1}
in the 6-31G basis with all electrons correlated.
For the size-intensivity tests, $n=0$--$4$ non-interacting helium
atoms are added at separations of $20$~\AA.
For these tests, the target relative residual was $10^{-9}$;
some calculations reached residuals of approximately $6\times10^{-8}$, which suffices to resolve the drifts in Table~\ref{tab:si}.  The uncontracted $U_2+U_3$ calculations use an explicit
determinant-basis triples space and the same Canonical reference gap.

\begin{table*}[t]
\caption{Comparison of CIS, CIS(D), and the three CIS(2) partitions
on QUEST\#1 (aug-cc-pVTZ). Errors are reported in eV relative to
the TBEs. The hydrogen chloride charge-transfer singlet contributes to the overall
statistics but not to the valence/Rydberg subsets.
CIS(D) values are those of Ref.~\onlinecite{Loos2018QUEST1}, except
for formaldehyde $2\,{}^3A_1$, for which our recalculated value of
7.58~eV replaces the published 7.68~eV.}\label{tab:quest}
\begin{ruledtabular}
% generated by gen_tables.py from quest1_cis2_singlet_triplet.csv (+ QUEST CIS(D))
\begin{tabular}{lccccc}
 & CIS & CIS(D) & Canonical & Block-diag. & Hybrid \\
\hline
\multicolumn{6}{l}{Singlets (53 states: 25 valence, 27 Rydberg)} \\
ME & $+0.52$ & $+0.07$ & $-0.30$ & $+0.29$ & $-0.01$ \\
MAE & 0.73 & 0.28 & 0.34 & 0.40 & 0.24 \\
MAE (valence) & 0.53 & 0.21 & 0.45 & 0.31 & 0.30 \\
MAE (Rydberg) & 0.92 & 0.36 & 0.24 & 0.49 & 0.20 \\
Max.\ abs.\ error & 1.95 & 1.06 & 1.12 & 1.37 & 1.03 \\
\hline
\multicolumn{6}{l}{Triplets (47 states: 29 valence, 18 Rydberg)} \\
ME & $-0.20$ & $+0.11$ & $-0.28$ & $+0.34$ & $-0.03$ \\
MAE & 0.61 & 0.22 & 0.29 & 0.37 & 0.15 \\
MAE (valence) & 0.64 & 0.20 & 0.35 & 0.37 & 0.17 \\
MAE (Rydberg) & 0.55 & 0.24 & 0.19 & 0.36 & 0.12 \\
Max.\ abs.\ error & 1.64 & 0.63 & 0.80 & 0.91 & 0.62 \\
\end{tabular}
\end{ruledtabular}
\end{table*}

\begin{figure*}[t]
\includegraphics[width=0.95\textwidth]{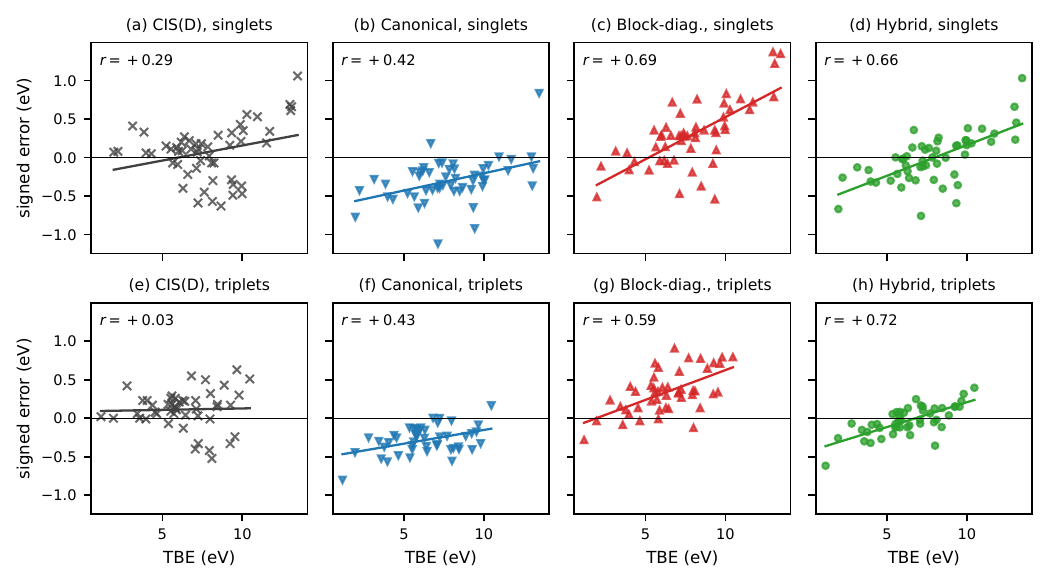}
\caption{Signed errors of CIS(D) and the three CIS(2) partitions against the
QUEST\#1 theoretical best estimates (aug-cc-pVTZ), for the 53 singlet
[panels (a)--(d)] and the 47 triplet [panels (e)--(h)] states.  Solid
lines are linear fits, with the Pearson correlation coefficient $r$
given in each panel; the slopes are listed in the Supporting
Information.}
\label{fig:quest}
\end{figure*}

\section{Results and discussion}
\label{sec:results}
\subsection{Comparison with ESMP2 on the sixteen-state benchmark}
\label{sec:esmp2_bench}

A comparison with ESMP2 places CIS(2) in the context of perturbation
theories built on an excited-state reference.  The sixteen singlet excitations studied by Shea, Gwin, and
Neuscamman~\cite{SheaGwinNeuscamman2020} provide published ESMP2 results
and common reference energies for this purpose.  This comparison assesses
the accuracy of the complete methods: ESMP2 uses an orbital-optimized
ESMF reference, whereas the present CIS(2) calculations use HF orbitals.
It therefore does not isolate the effects of triples contraction or
partitioning from those of reference relaxation.

Table~\ref{tab:table2} compares the errors in the cc-pVDZ basis against
$\delta$-CR-EOM-CC(2,3),D reference energies.  The ESMP2 statistics are
recomputed from the published errors for these sixteen states, excluding
the two intermolecular charge-transfer excitations in the original set.

Among the three CIS(2) partitions, Hybrid has the smallest MAE,
0.19~eV, compared with 0.24~eV for Canonical and 0.23~eV for
Block-diagonal.  Its MAE is also below those of CIS(D), EOM-CCSD,
and CC2 on this set (0.32, 0.22, and 0.33~eV, respectively).

The doubles--triples coupling (Sec.~\ref{sec:workeq}) is essential
for both Canonical and Hybrid.  In CIS(D), this coupling vanishes by
the Brillouin condition, but with the CIS-state Fock matrix it is part
of the zeroth-order Hamiltonian, and setting it to zero has no
justification within the present partition.  Indeed, doing so while
keeping all other quantities fixed raises the MAE from 0.24 to 0.45~eV
for Canonical and from 0.19 to 0.32~eV for Hybrid.

ESMP2 gives the lowest MAE among the tested methods, 0.12~eV.
Both ESMP2 and Hybrid have small mean errors ($-0.03$ and $-0.04$~eV,
respectively), but Hybrid has a larger maximum absolute error
(0.62 vs.\ 0.42~eV). Hybrid thus remains less accurate than ESMP2, a difference that
   may reflect the orbital relaxation included in the ESMF reference.

\begin{figure*}[t]
\includegraphics[width=0.95\textwidth]{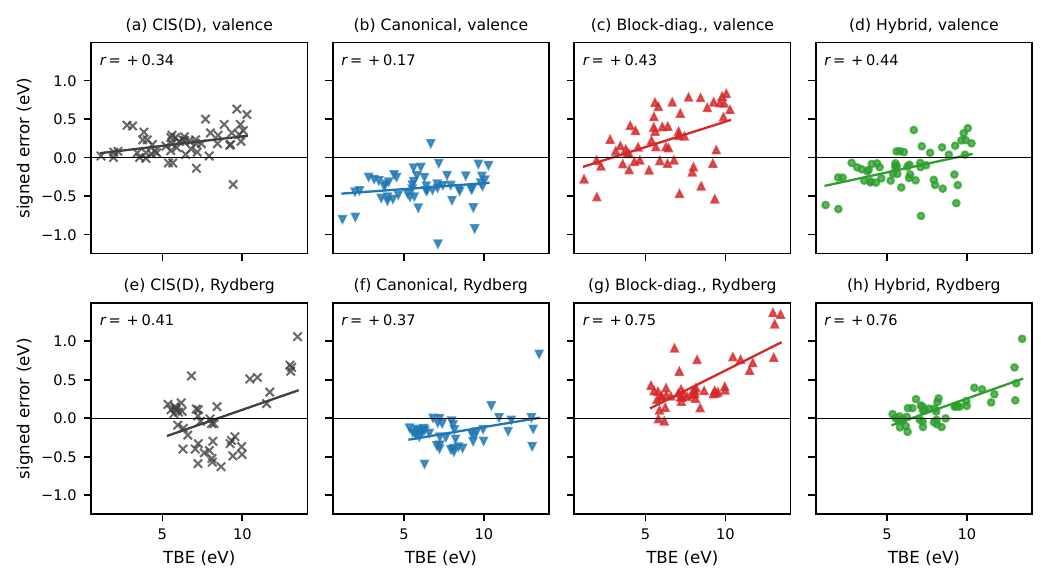}
\caption{Same data as Fig.~\ref{fig:quest}, grouped by state character:
the 54 valence [panels (a)--(d)] and the 45 Rydberg [panels (e)--(h)]
states of both spin manifolds, with linear fits as in
Fig.~\ref{fig:quest}.  The $1\,{}^1\Pi$ state of hydrogen chloride,
classified as charge transfer in QUEST\#1, belongs to neither group.}
\label{fig:quest_vr}
\end{figure*}

\subsection{QUEST\#1 singlet and triplet benchmark}
\label{sec:quest}

Next, we extend the benchmark to QUEST\#1, which covers a broader
range of states, including both singlet and triplet excitations
of valence and Rydberg character.
The molecules in the ESMP2 benchmark also occur in QUEST\#1,
so this is an expansion of state coverage rather than an
independent molecular test set. The aug-cc-pVTZ basis and QUEST theoretical best
estimates also differ from those used in the preceding comparison.  

Table~\ref{tab:quest} summarizes the errors relative to the QUEST\#1
theoretical best estimates.  State-resolved errors are shown in
Fig.~\ref{fig:quest}, and the individual excitation energies are
listed in the Supporting Information.

Canonical and Block-diagonal use the same reference gap in both
excitation sectors, but their mean errors have opposite signs.
Canonical underestimates the singlet and triplet excitation energies
by 0.30 and 0.28~eV on average, whereas Block-diagonal overestimates
them by 0.29 and 0.34~eV. This ordering is consistent with the
smaller Canonical denominators and the larger negative correlation
correction they produce.  CIS(D) has a smaller positive bias, with
mean errors of $+0.07$ and $+0.11$~eV.

Figure~\ref{fig:quest} shows that these biases are shared by many
states rather than being dominated by a few outliers. Canonical gives
negative errors for 49 of the 53 singlets and 44 of the 47 triplets,
whereas Block-diagonal gives positive errors for 39 singlets and
41 triplets.
Each panel also includes a linear fit of the signed errors against
the reference excitation energy and the corresponding Pearson
correlation coefficient $r$. The slopes are listed in the
Supporting Information.

Hybrid changes the doubles reference gap to $\wcis$ while retaining
$\wzero$ in the contracted-triples sector. In Fig.~\ref{fig:quest},
its excitation energy lies between the Canonical and Block-diagonal
values for every state in both spin manifolds. The separation varies
from state to state, reflecting the different responses of the coupled
amplitudes to the partition choice. Its overall mean errors
are close to zero ($-0.01$~eV for singlets and $-0.03$~eV for
triplets), and its mean absolute errors of 0.24 and 0.15~eV are the
lowest among the methods in Table~\ref{tab:quest}.
Hybrid also gives the lowest MAE among the three CIS(2) partitions
in each of the four subsets: valence singlets, Rydberg singlets,
valence triplets, and Rydberg triplets. The Hybrid triplet errors have a narrower distribution than
the singlet errors in Fig.~\ref{fig:quest}.

\begin{table*}[t]
\caption{Contracted CIS(2) versus the uncontracted $U_2+U_3$ first-order space
(Canonical partition, 6-31G, all electrons; eV).
$\delta=\omega_{U_3}-\omega_{\rm Can}$.}
\label{tab:u3}
\begin{ruledtabular}
% generated by gen_tables.py: contracted vs uncontracted at n = 0 (eV); delta = U3 - Canonical
\begin{tabular}{llcccc}
molecule & state & CIS & Canonical & Uncontracted & $\delta$ \\
\hline
formaldehyde & $S_{1}$ & 4.297 & 3.651 & 3.628 & $-0.023$ \\
formaldehyde & $S_{2}$ & 9.423 & 9.021 & 9.006 & $-0.014$ \\
formaldehyde & $S_{3}$ & 10.257 & 10.137 & 10.109 & $-0.028$ \\
formaldehyde & $S_{4}$ & 11.705 & 10.639 & 10.615 & $-0.024$ \\
formaldehyde & $S_{5}$ & 11.814 & 9.342 & 9.277 & $-0.065$ \\
formaldehyde & $S_{6}$ & 13.076 & 11.431 & 11.387 & $-0.044$ \\
water & $S_{1}$ & 9.405 & 8.368 & 8.362 & $-0.006$ \\
water & $S_{2}$ & 11.337 & 10.621 & 10.615 & $-0.006$ \\
water & $S_{3}$ & 11.858 & 10.707 & 10.704 & $-0.003$ \\
water & $S_{4}$ & 13.936 & 13.219 & 13.214 & $-0.005$ \\
water & $S_{5}$ & 15.497 & 15.268 & 15.265 & $-0.004$ \\
water & $S_{6}$ & 19.328 & 18.753 & 18.740 & $-0.013$ \\
\end{tabular}
\end{ruledtabular}
\end{table*}

Grouping the states by valence or Rydberg character in
Fig.~\ref{fig:quest_vr} distinguishes where Hybrid improves on CIS(D)
and where sizable errors remain. For valence excitations, CIS(D)
shows a systematic tendency to overestimate the excitation energies,
with positive errors for 47 of the 54 states and a mean error of
$+0.18$~eV. Hybrid instead underestimates 39 states, with a mean
error of $-0.14$~eV. Their valence MAEs are comparable (0.21~eV for
CIS(D) and 0.23~eV for Hybrid), although CIS(D) is more accurate
for the singlet subset (0.21 versus 0.30~eV). The largest valence underestimations by Hybrid occur for the singlet $\pi\rightarrow\pi^*$ excitation of
streptocyanine-C1 ($-0.76$~eV) and the singlet and triplet
$n\rightarrow\pi^*$ excitations of nitrosomethane ($-0.67$ and
$-0.62$~eV). All three CIS(2) partitions underestimate these states,
so their errors are not removed by changing the partition.

For Rydberg excitations, CIS(D) no longer shows the predominantly
positive errors found for valence states.
Its errors are distributed almost equally between positive and
negative values (22 and 23 states, respectively), with substantial
scatter even at similar excitation energies. The near-zero mean
error of $-0.02$~eV therefore reflects cancellation rather than
uniform accuracy: the MAE is 0.31~eV. Hybrid gives more accurate
Rydberg excitation energies, with an MAE of 0.17~eV. This improvement occurs in both spin manifolds. The singlet and
triplet Rydberg MAEs decrease from 0.36 and 0.24~eV with CIS(D)
to 0.20 and 0.12~eV with Hybrid, respectively. Despite this
improvement, Hybrid retains a tendency to overestimate Rydberg
excitation energies, with a mean error of $+0.13$~eV.
The largest overestimation occurs for the N$_2$ Rydberg excitation
with a TBE of 13.46~eV, for which the Hybrid error is $+1.03$~eV.
All three partitions overestimate this excitation.

We attribute the improved Rydberg performance to the state-specific Fock operator
of CIS(2), which reflects the changes in the mean field and screening
associated with excitation. These changes are expected to be
particularly important for Rydberg states, where the excited electron
occupies a diffuse orbital and the density differs markedly from
that of the ground state. CIS(D), by contrast, retains the
ground-state HF mean field.

In Figs.~\ref{fig:quest} and \ref{fig:quest_vr}, the CIS(2) errors show a clear tendency to increase with the reference
excitation energy, particularly for Block-diagonal and Hybrid.
Over the 100 states, the correlation coefficients are
$r=0.39$, $0.61$, and $0.65$ for Canonical, Block-diagonal, and
Hybrid, respectively, compared with $0.18$ for CIS(D).
For Hybrid, the fitted slope is $+0.069$~eV per eV, and the
correlation remains strong within the Rydberg subset ($r=0.76$),
showing that the overall trend is not solely due to differences
between valence and Rydberg states.
The slope remains positive in all four spin/character subsets
for both Hybrid and Block-diagonal.
For CIS(D), the trend is less consistent across subsets.
Its correlation coefficients are $0.59$ for Rydberg singlets
and $0.57$ for valence triplets, but $-0.15$ for Rydberg triplets.
The systematic component of the CIS(2) errors suggests a route
to further improvement through modifications of the external
denominators, which we investigate in subsequent work.

\begin{table*}[t]
\caption{Size-intensivity test: $\omega(n{=}4)-\omega(n{=}0)$ in meV for water plus
$n$ non-interacting helium atoms (6-31G, all electrons, GMRES threshold
$10^{-9}$).  $S_1$--$S_6$ denote the six lowest singlet CIS states of water.}
\label{tab:si}
\begin{ruledtabular}
% generated by gen_tables.py from he_si_hf_water_root*.csv (+ he_si_hybproj_water_root*.csv) (drift = omega(n_max) - omega(0), meV)
\begin{tabular}{lcccccc}
state & CIS(D) & Canonical (unproj.) & Canonical (proj.) & Block-diag. & Hybrid & Uncontracted $U_2{+}U_3$ \\
\hline
$S_{1}$ & 0.000 & 0.000 & 0.000 & +36.313 & 0.000 & -0.033 \\
$S_{2}$ & 0.000 & 0.000 & 0.000 & +33.546 & 0.000 & -0.037 \\
$S_{3}$ & 0.000 & -0.015 & 0.000 & +42.049 & 0.000 & -0.817 \\
$S_{4}$ & 0.000 & 0.000 & 0.000 & +33.477 & 0.000 & -0.417 \\
$S_{5}$ & 0.000 & 0.000 & 0.000 & +23.592 & 0.000 & -0.407 \\
$S_{6}$ & 0.000 & -0.051 & 0.000 & +30.440 & 0.000 & -2.329 \\
\end{tabular}
\end{ruledtabular}
\end{table*}

\subsection{Validity of the contracted triples}
\label{sec:t2c1_valid}

As established in Sec.~\ref{sec:h0}, the contracted triples space
contains all triples components generated by general one- and
two-body operators acting on $|S\rangle$. In particular, it
contains the triples component of $\hat Q\Ham|S\rangle$ exactly.
Triples orthogonal to this space have no direct Hamiltonian coupling
to the CIS reference, but may contribute indirectly through the
zeroth-order operator. We assess the effect of omitting this response
by comparing the contracted and uncontracted triples spaces while
retaining the full doubles space, the same CIS reference, and the
Canonical partition in both calculations.

Table~\ref{tab:u3} gives the excitation energies
for the six lowest singlet states of water and formaldehyde in the
6-31G basis. The uncontracted energies are lower in all twelve cases,
by up to 0.013~eV for water and 0.065~eV for formaldehyde.
Thus, allowing the triples response to extend beyond
the internally contracted space changes the excitation energies
only modestly for the systems tested here.

Each contracted triple is a fixed linear combination of uncontracted
triples,
$\tmu|S\rangle=\sum_{ai}c_{ai}\tmu|\Phi_i^a\rangle$,
so the contracted space is a subspace of $U_2+U_3$.
For a positive-definite first-order operator $\hat H_0-E^{(0)}$,
the observed ordering follows from the Hylleraas variational principle,
since restricting the first-order space can only raise the
second-order energy. The same ordering then holds for the excitation
energies because the ground-state MP2 correction is identical in both
calculations.

\subsection{Size-intensivity}
\label{sec:si_test}

In this section, we test the spectator cancellation derived in
Sec.~\ref{sec:sizeintensive}.
Table~\ref{tab:si} lists, for the six lowest singlet states of water, the change
of the excitation energy when four non-interacting helium atoms are added.
CIS(D) and the projected contracted Canonical and Hybrid partitions are
size-intensive to the solver tolerance.  The Block-diagonal partition drifts
by tens of millielectronvolts, linearly in the number of spectators, as the
chromophore-dependent denominator shift $\wzero-\wcis$ of
Sec.~\ref{sec:sizeintensive} predicts.  The uncontracted $U_2+U_3$ space
drifts as well, by $0.008$--$0.582$~meV per helium atom and again exactly linearly in
$n$.  Without the Fock projection of Eq.~\eqref{eq:fperp} the two totally
symmetric states ($A_1$; $S_3$ and $S_6$) drift by $0.004$ and $0.013$~meV per
helium atom in the Canonical partition, with nearly identical drifts in
Hybrid.  This drift is the
residual coupling of the spectator triple to the spectator double through the
occupied--virtual Fock block discussed in Sec.~\ref{sec:sizeintensive}.  The
projection removes it identically (below $10^{-4}$~meV over four atoms)
while shifting the $n=0$ energies by at most $4.1$~meV.  The other four states are
unaffected by the projection.  For the pure-singles references considered here, the projected Canonical
and Hybrid results therefore satisfy size-intensivity within the solver
tolerance.

\section{Conclusions}
\label{sec:conclusions}

We have developed CIS(2), a state-specific second-order correction
defined by a CIS reference and its generalized Fock operator.
The first-order space follows from a fully internally
contracted construction for a full-rank CIS coefficient matrix and is
extended to retain all doubles when the matrix is rank deficient.
It contains $\hat Q\Ham|S\rangle$ exactly, while restricting the
triples response to the internally contracted space. Spin adaptation and
semicanonical orbitals give a matrix-free amplitude problem with
$O(o^3v^2)$ cost per iteration and spatial tensors of dimension
$o^2v^2$.

The contraction also plays a role in size-intensivity,
whose preservation requires an appropriate partition.
Using the bare Fock gap in the contracted-triples sector, together
with projection of the residual occupied--virtual Fock coupling,
recovers the spectator MP2 correlation energy and ensures its
exact cancellation from the excitation energy.
Neither the Block-diagonal partition nor the uncontracted
Canonical variant satisfies these conditions.
The doubles reference gap remains unconstrained by this cancellation,
allowing the Hybrid partition to combine the CIS excitation energy
in the doubles sector with the bare Fock gap in the triples sector.

The benchmarks favor this Hybrid choice over the Canonical
and Block-diagonal partitions. On QUEST\#1, Hybrid gave the lowest
MAE among the three partitions for all four combinations of spin
multiplicity and valence/Rydberg character. It reduced the substantial
negative bias of Canonical and achieved better overall accuracy than
CIS(D). The improvement over CIS(D) was largest for Rydberg states,
whereas CIS(D) remained more accurate for valence singlets.
On the sixteen-state singlet benchmark, Hybrid was competitive
with EOM-CCSD, although orbital-relaxed ESMP2 remained more accurate.

The systematic energy dependence of the residual errors suggests
scope for further improvement through modifications of the external
denominators, which we investigate in forthcoming work.
The present HF-orbital formulation also provides a foundation
for correlation corrections to orbital-optimized and state-averaged
CIS references. These extensions are currently underway.

\begin{acknowledgments}
This work was supported by the JST FOREST Program, Grant No. JPMJFR223U.
\end{acknowledgments}

\section*{Associated Content}
\subsection*{Supporting Information}
The fully internally contracted space generated from a CIS reference;
explicit semicanonical spin-orbital and spin-adapted working equations
with packed amplitude storage; and state-resolved QUEST\#1 excitation
energies with the correlation coefficients and slopes of the fits of
the errors against the reference excitation energies.

\section*{Data availability}
The numerical results supporting the findings of this
study are provided in the article and its Supporting Information.
Additional data and benchmark input files are available from the
author upon reasonable request.

\bibliography{refs}

\end{document}